\PassOptionsToPackage{unicode}{hyperref}
\PassOptionsToPackage{hyphens}{url}
\documentclass[
]{article}
\usepackage{amsmath,amssymb}
\usepackage{lmodern}
\usepackage{ifxetex,ifluatex}
\ifnum 0\ifxetex 1\fi\ifluatex 1\fi=0 
  \usepackage[T1]{fontenc}
  \usepackage[utf8]{inputenc}
  \usepackage{textcomp} 
\else 
  \usepackage{unicode-math}
  \defaultfontfeatures{Scale=MatchLowercase}
  \defaultfontfeatures[\rmfamily]{Ligatures=TeX,Scale=1}
\fi
\IfFileExists{upquote.sty}{\usepackage{upquote}}{}
\IfFileExists{microtype.sty}{
  \usepackage[]{microtype}
  \UseMicrotypeSet[protrusion]{basicmath} 
}{}
\makeatletter
\@ifundefined{KOMAClassName}{
  \IfFileExists{parskip.sty}{%
    \usepackage{parskip}
  }{
    \setlength{\parindent}{0pt}
    \setlength{\parskip}{6pt plus 2pt minus 1pt}}
}{
  \KOMAoptions{parskip=half}}
\makeatother
\usepackage{xcolor}
\IfFileExists{xurl.sty}{\usepackage{xurl}}{} 
\IfFileExists{bookmark.sty}{\usepackage{bookmark}}{\usepackage{hyperref}}
\hypersetup{
  pdftitle={Reproducibility of COVID-19 pre-prints},
  pdfauthor={Annie Collins; Rohan Alexander},
  hidelinks,
  pdfcreator={LaTeX via pandoc}}
\usepackage[margin=1in]{geometry}
\usepackage{longtable,booktabs,array}
\usepackage{calc} 
\usepackage{etoolbox}
\makeatletter
\patchcmd\longtable{\par}{\if@noskipsec\mbox{}\fi\par}{}{}
\makeatother
\IfFileExists{footnotehyper.sty}{\usepackage{footnotehyper}}{\usepackage{footnote}}
\makesavenoteenv{longtable}
\usepackage{graphicx}
\makeatletter
\def\maxwidth{\ifdim\Gin@nat@width>\linewidth\linewidth\else\Gin@nat@width\fi}
\def\maxheight{\ifdim\Gin@nat@height>\textheight\textheight\else\Gin@nat@height\fi}
\makeatother
\setkeys{Gin}{width=\maxwidth,height=\maxheight,keepaspectratio}
\makeatletter
\def\fps@figure{htbp}
\makeatother
\usepackage{booktabs}
\usepackage{longtable}
\usepackage{array}
\usepackage{multirow}
\usepackage{wrapfig}
\usepackage{float}
\usepackage{colortbl}
\usepackage{pdflscape}
\usepackage{tabu}
\usepackage{threeparttable}
\usepackage{threeparttablex}
\usepackage[normalem]{ulem}
\usepackage{makecell}
\usepackage{xcolor}
\ifluatex
  \usepackage{selnolig}  
\fi
\newlength{\cslhangindent}
\newlength{\csllabelwidth}
\newenvironment{CSLReferences}[2] 
 {
  \setlength{\parindent}{0pt}
  \ifodd #1 \everypar{\setlength{\hangindent}{\cslhangindent}}\ignorespaces\fi
  \ifnum #2 > 0
  \setlength{\parskip}{#2\baselineskip}
  \fi
 }%
 {}
\usepackage{calc}

\title{Reproducibility of COVID-19 pre-prints\thanks{\textbf{Preprint:} This present manuscript is a version of a preprint by the authors (Collins and Alexander 2021). The text, tables and figures of this manuscript overlap with those of the preprint version by the authors. \textbf{Acknowledgements:} Two anonymous reviewers and the editor provided extensive comments that substantially improved this paper and we are grateful to them. We thank Amy Farrow, Jessica Gronsbell, Monica Alexander, and Thomas William Rosenthal for helpful comments. \textbf{Funding statement:} Collins thanks CANSSI Ontario and Monica Alexander for financial support. \textbf{Data accessibility statement:} Code and data are available at: \url{https://github.com/anniecollins/reproducibility_markers_in_covid19_preprints}. \textbf{Author contributions:} Collins had the original idea and wrote the first draft of the paper. Collins and Alexander obtained and created the datasets. Collins and Alexander analyzed the data. All authors interpreted the data and contributed writing the paper and approved the final version.}}
\author{Annie Collins\footnote{University of Toronto.} \and Rohan Alexander\footnote{University of Toronto, \href{mailto:rohan.alexander@utoronto.ca}{\nolinkurl{rohan.alexander@utoronto.ca}}.}}
\date{13 January 2022}

\begin{document}
\maketitle
\begin{abstract}
To examine the reproducibility of COVID-19 research, we create a dataset of pre-prints posted to arXiv, bioRxiv, and medRxiv between 28 January 2020 and 30 June 2021 that are related to COVID-19. We extract the text from these pre-prints and parse them looking for keyword markers signaling the availability of the data and code underpinning the pre-print. For the pre-prints that are in our sample, we are unable to find markers of either open data or open code for 75 per cent of those on arXiv, 67 per cent of those on bioRxiv, and 79 per cent of those on medRxiv.
\end{abstract}

\hypertarget{introduction}{%
\section{Introduction}\label{introduction}}

Scientists use open repositories of papers to disseminate their research more quickly than is possible in traditional journals or conference proceedings, and to obtain feedback on their work prior to publication. These repositories, such as arXiv, bioRxiv, and medRxiv, are a critical component of scientific communication and a lot of research builds on the pre-prints posted there. Pre-print repositories have been especially important during the 2019 novel coronavirus (COVID-19) pandemic and the changes it has imposed on the scientific community (Else 2020). The centrality of pre-prints to science means that it is important that the results that are posted are credible. These repositories are not peer-reviewed, and, in general, anyone with appropriate academic credentials can submit a pre-print.

Neither peer-review nor credentials are a panacea nor a guarantee of quality. And the gate-keeping and slow publication times of traditional journals mean pre-print repositories are important. But it is important that scientists impose standards on themselves, and arguably repositories have a role to play here. Following Weissgerber et al. (2021), we examine pre-prints about COVID-19 posted to arXiv, bioRxiv, and medRxiv from 28 January 2020 through to 30 June 2021. By way of background, each of these three repositories has a different focus: arXiv is general although it has especially high rates of usage from fields like mathematics, physics, and computer science, bioRxiv focuses on biological sciences, and medRxiv focuses on health sciences.

We search for markers of open science as indicators of reproducibility, specifically open data and open code. The definition of reproducibility tends to vary by context and academic field (Barba 2018). For the purposes of this paper, we define reproducibility to mean the ability for different researchers to achieve the same results given the same data and computational methods as the original source. This contrasts with replicability, which we define as the ability for different researchers to achieve consistent results by conducting the full data collection and analysis process in lieu of reusing original data. These definitions match that of National Academies of Sciences and Medicine (2019) and Cacioppo et al. (2015). What constitutes open code or open data is complicated and discipline specific. The details of the \texttt{oddpub} approach are available in Riedel, Kip, and Bobrov (2020). The general criteria are that specific mention should be made of where the data and code are located, and that data should be as close to raw as possible. Data and code must also be freely accessible to anyone (no request, application, registration process, or affiliation required).

\begin{figure}

{\centering \includegraphics[width=0.9\linewidth]{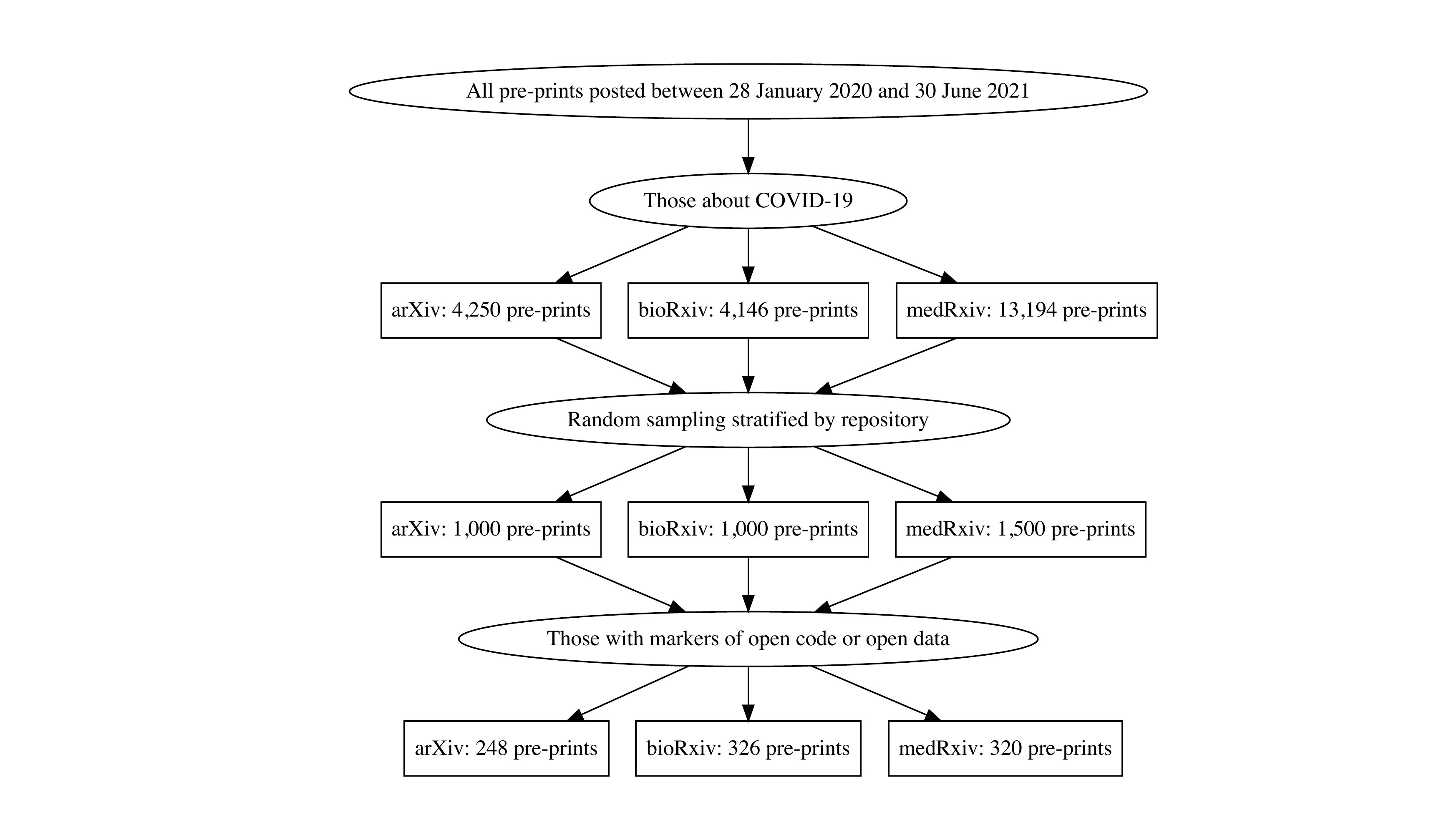} 

}

\caption{Summary of process and main results}\label{fig:flowchart}
\end{figure}

We find that of the papers sampled, approximately 75 per cent of papers from arXiv, 67 per cent of papers from bioRxiv, and 79 per cent of papers from medRxiv contain neither open data nor open code markers. A summary of our main results is contained in Figure \ref{fig:flowchart}. Examining trends over time, we find that the proportion of pre-prints containing open data or code markers has fluctuated but shown no obvious trend throughout the pandemic. We also find that the presence of open data or open code markers seems to have little association with a pre-print's subsequent publication, and the subset of sampled pre-prints that have been published contains approximately the same proportion of papers with these markers.

The remainder of this paper is structured as follows: in Section \ref{methodology} we discuss the process of constructing our dataset through retrieving pre-prints from the arXiv, bioRxiv, and medRxiv repositories and mining them for open data and open code markers. In Section \ref{results}, we present the results and key findings of this process. Finally, in Section \ref{discussion} we discuss the implications of these findings in the broader context of reproducibility and science during the COVID-19 pandemic, as well as next steps to expand on our findings and questions raised in the research process.

\hypertarget{methodology}{%
\section{Methodology}\label{methodology}}

\hypertarget{pre-print-metadata}{%
\subsection{Pre-print metadata}\label{pre-print-metadata}}

Our primary dataset consists of pre-print metadata extracted from the arXiv, bioRxiv, and medRxiv pre-print repositories via their respective Application Programming Interfaces (APIs). This metadata varies by repository, but generally includes: title, abstract, author(s), date created, research field, DOI, version number, corresponding author, corresponding author's institutional affiliation, published DOI (if the pre-print has since been published in a peer-reviewed journal), and download link. The data collection process was conducted separately for COVID-19 and pre-COVID-19 papers.

For COVID-19-related pre-prints, we first created a local copy of each repository containing all metadata for pre-prints posted between 1 January, 2020, and 30 June, 2021.
We classified individual pre-prints as ``COVID-19-related'' based on whether they contained one or more of the following terms in their title or abstract (case insensitive): ``COVID-19,'' ``COVID 19,'' ``corona virus,'' ``coronavirus,'' ``coronavirus-2,'' ``SARS-CoV-2,'' ``SARSCoV-2,'' and ``2019-nCoV.'' We then randomly sampled pre-prints for further analysis.

For pre-COVID-19 pre-prints, we created a local copy of each repository containing all metadata for pre-prints posted between 1 January, 2019, and December 31, 2019. Since medRxiv was launched in June 2019, we used all pre-print data from the latter half of 2019. We then randomly sampled 1,200 pre-prints from each repository's dataset for analysis, except for medRxiv for which only 913 pre-prints were available over this time.

\hypertarget{open-data-and-code-detection}{%
\subsection{Open data and code detection}\label{open-data-and-code-detection}}

We checked our sampled pre-prints for open data and code markers using the Open Data Detection in Publications (ODDPub) text mining algorithm (Riedel, Kip, and Bobrov 2020) within the \texttt{oddpub} R package (Riedel 2019) (RRID:SCR\_018385) . This required downloading each pre-print as a PDF and then converting the PDFs to text files.
We then conducted the open data and open code detection procedure, which involved searching for keywords and other markers of open data and open code availability. This was conducted using the \texttt{open\_data\_search()} function from the \texttt{oddpub} package. In the validation conducted by the authors of the package, the ODDPub algorithm had a sensitivity of 0.73 and a specificity of 1.00 for open code detection, and a sensitivity of 0.73 and a specificity of 0.97 for open data detection compared with manual screening (Riedel, Kip, and Bobrov 2020). Since the ODDPub algorithm was developed specifically for biomedical publications, we conducted our own validation process for its performance on arXiv pre-prints. We found that the ODDPub algorithm performed with a sensitivity of 0.60 and a specificity of 0.98 for open code detection, and a sensitivity of 0.67 and a specificity of 0.98 for open data detection compared with manual screening. Details of our validation procedure are contained in Appendix \ref{oddpub-algorithm-performance-on-arxiv-pre-prints}. Our work was conducted using the statistical programming language R (R Core Team 2020) (RRID:SCR\_001905).

The result of this process is a dataset indicating the presence of open data or open code markers in each pre-print (with a logical vector for each marker, followed by the relevant open data or open code statements where applicable). Our final dataset was formed by joining this output with the original sample metadata, typically using the DOI or the unique file name, to form a dataset including all original metadata for each pre-print alongside its open data and open code status and markers.

\hypertarget{results}{%
\section{Results}\label{results}}

\hypertarget{pre-pandemic-pre-prints}{%
\subsection{Pre-pandemic pre-prints}\label{pre-pandemic-pre-prints}}

To examine the influence of the COVID-19 pandemic on open science practices during the pandemic, we analyzed pre-prints posted between January and December 2019 from each of the four repositories in question. Since medRxiv was founded in June 2019, all pre-prints posted in the latter half of 2019 were analyzed (a total of 913). For all other repositories, a random sample of 1,200 was taken from all non-COVID-19-related pre-prints posted in the relevant date range.

\begin{figure}[H]

{\centering \includegraphics[width=0.9\linewidth]{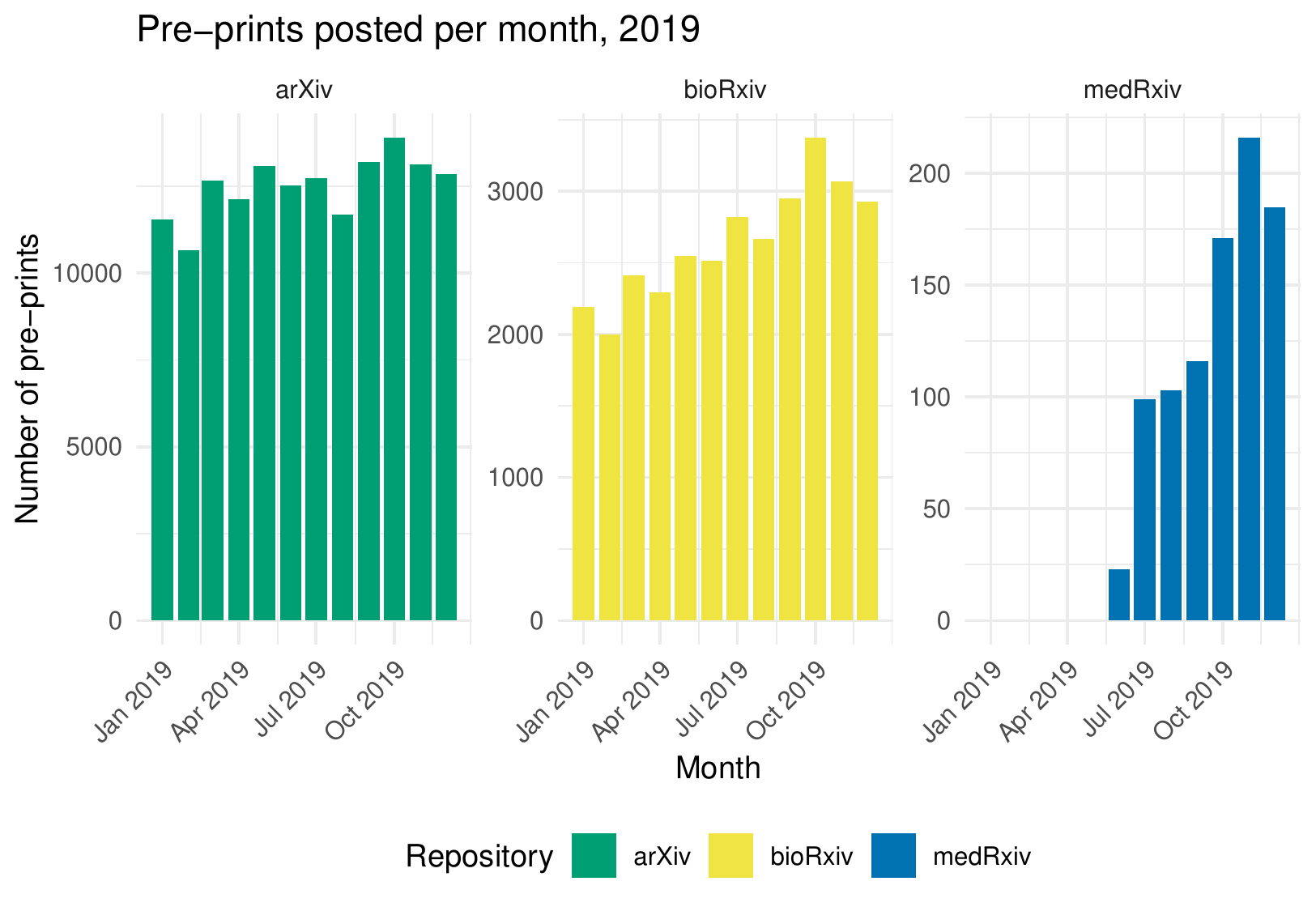} 

}

\caption{Number of pre-prints posted per month in 2019 to arXiv (N = 150,018), bioRxiv (N = 31,752), medRxiv (N = 913)}\label{fig:2019-results-med-arxiv}
\end{figure}

Between June and December 2019, the number of pre-prints posted to medRxiv monthly saw an overall increase, which may be expected as the repository gained recognition and popularity in the medical research community (Figure \ref{fig:2019-results-med-arxiv}). The number of pre-prints posted monthly to bioRxiv also saw a slight overall increase throughout 2019, while the number of those posted to arXiv fluctuated throughout the year (Figure \ref{fig:2019-results-med-arxiv}). Due to its relative immaturity at the beginning of the COVID-19 pandemic, a significant portion of medRxiv's overall usage has been dedicated to COVID-19-related research. In total, 21,647 pre-prints were posted to medRxiv between June 2019 and 30 June, 2021, 13,194 of which (approximately 61 per cent) relate to COVID-19.

Of the analyzed pre-prints from 2019, 93 per cent of those posted to arXiv, 63 per cent of those posted to bioRxiv, and 75 per cent of those posted to medRxiv showed no indication of open data or open code.

\begin{table}

\caption{\label{tab:2019-published}Counts and proportion of published COVID-19 pre-prints in each repository, 2019}
\centering
\begin{tabular}[t]{lrr}
\toprule
Repositiory & Published pre-prints & Proportion published\\
\midrule
arXiv & 62,045 & 0.41\\
bioRxiv & 20,445 & 0.64\\
medRxiv & 555 & 0.61\\
\bottomrule
\end{tabular}
\end{table}

\begin{table}

\caption{\label{tab:2019-open-publishing}Counts and proportions of open data markers by whether the pre-print was published, 2019}
\centering
\begin{tabular}[t]{lrrrrr}
\toprule
Status & Both & Neither & Open code & Open data & Proportion with neither\\
\midrule
\addlinespace[0.3em]
\multicolumn{6}{l}{\textbf{arXiv}}\\
\hspace{1em}Unpublished & 11 & 645 & 44 & 3 & 0.92\\
\hspace{1em}Published & 6 & 467 & 15 & 9 & 0.94\\
\addlinespace[0.3em]
\multicolumn{6}{l}{\textbf{bioRxiv}}\\
\hspace{1em}Unpublished & 38 & 264 & 34 & 87 & 0.62\\
\hspace{1em}Published & 78 & 488 & 64 & 147 & 0.63\\
\addlinespace[0.3em]
\multicolumn{6}{l}{\textbf{medRxiv}}\\
\hspace{1em}Unpublished & 19 & 277 & 19 & 43 & 0.77\\
\hspace{1em}Published & 38 & 409 & 36 & 72 & 0.74\\
\bottomrule
\end{tabular}
\end{table}

Examining publication rates for pre-pandemic papers, we observe that 41 per cent of pre-prints posted to arXiv, 64 per cent of pre-prints posted to bioRxiv, and 61 per cent of pre-prints posted to medRxiv during 2019 were eventually peer reviewed and published (Table \ref{tab:2019-published}). When disaggregated by open data and code status, we find that published and unpublished pre-prints contain open data and code markers in similar proportions (Table \ref{tab:2019-open-publishing}).

\hypertarget{all-pre-prints-related-to-covid-19}{%
\subsection{All pre-prints related to COVID-19}\label{all-pre-prints-related-to-covid-19}}

\begin{figure}[H]

{\centering \includegraphics[width=0.9\linewidth]{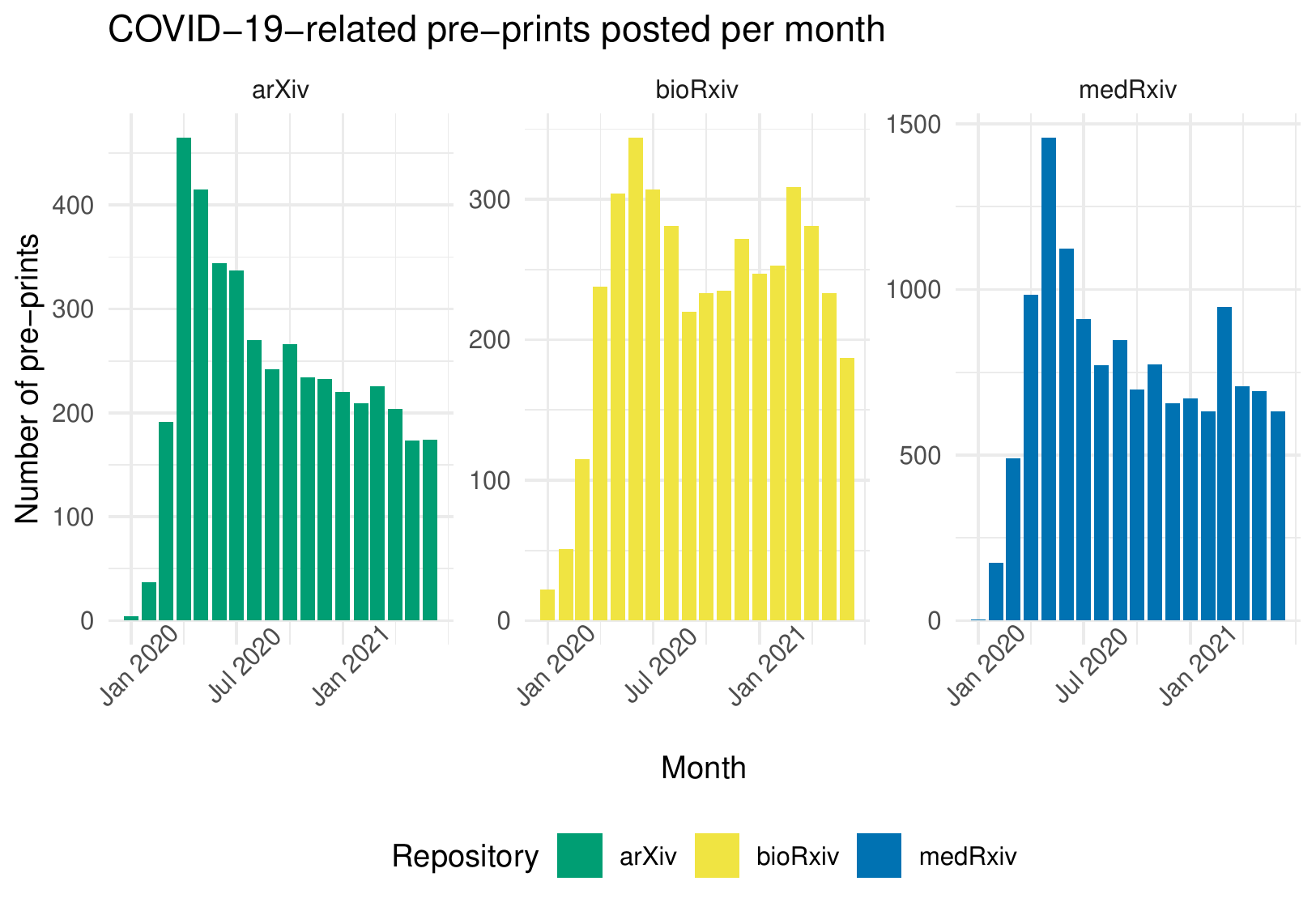} 

}

\caption{Number of pre-prints related to COVID-19 posted to arXiv (N = 4,250), bioRxiv (N = 4,146), and medRxiv (N = 13,194) per month}\label{fig:monthly-papers-total}
\end{figure}

The number of pre-prints posted per month increased in the first half of 2020 across all repositories, reaching a maximum sometime between April and June (depending on repository) and subsequently decreasing. The number of pre-prints posted monthly since August 2020 has remained reasonably steady, with the exception of medRxiv, which experienced an increase to nearly 1,000 pre-prints posted in March 2021 (Figure \ref{fig:monthly-papers-total}). For context, COVID-19 was declared a pandemic by the World Health Organization (WHO) on March 11, 2020, at which point the number of cases globally had just surpassed 118,000 (primarily in east Asia) and the virus had been reported in 114 countries (World Health Organization 2020).

\hypertarget{open-data-and-code}{%
\subsection{Open data and code}\label{open-data-and-code}}

From the collection of all pre-prints related to COVID-19, we randomly sampled 3,500 pre-prints to analyze, stratified by repository. This sample is broken down as follows: 1,500 from medRxiv, 1,000 from arXiv, and 1,000 from bioRxiv. Broadly, we are unable to find markers of either open data or open code for 2,606 pre-prints or approximately 74 per cent of our sample (Appendix \ref{supporting-tables} Table \ref{tab:summarycounts-repositories}). Of the remaining pre-prints, 7 per cent contained open code markers only, 10 per cent contained open data markers only, and 8 per cent included markers of both open data and open code.

When differentiated by repository, we observe that open data and code markers were absent from 75 per cent of the sampled arXiv pre-prints, 67 per cent of the sampled bioRxiv pre-prints, 79 per cent of the sampled medRxiv pre-prints. The distribution of the remaining portion of pre-prints also varies by repository (Appendix \ref{supporting-tables} Table \ref{tab:summarycounts-repositories}). Notably, 28 per cent of sampled pre-prints from bioRxiv contained open data markers and 22 per cent of sampled arXiv pre-prints contained markers of open code, the highest proportions of any repository for each type of marker. Our results are similar to McGuinness and Sheppard (2021), who focus on medRxiv and find that 23 per cent describe open data.

\begin{figure}

{\centering \includegraphics[width=0.9\linewidth]{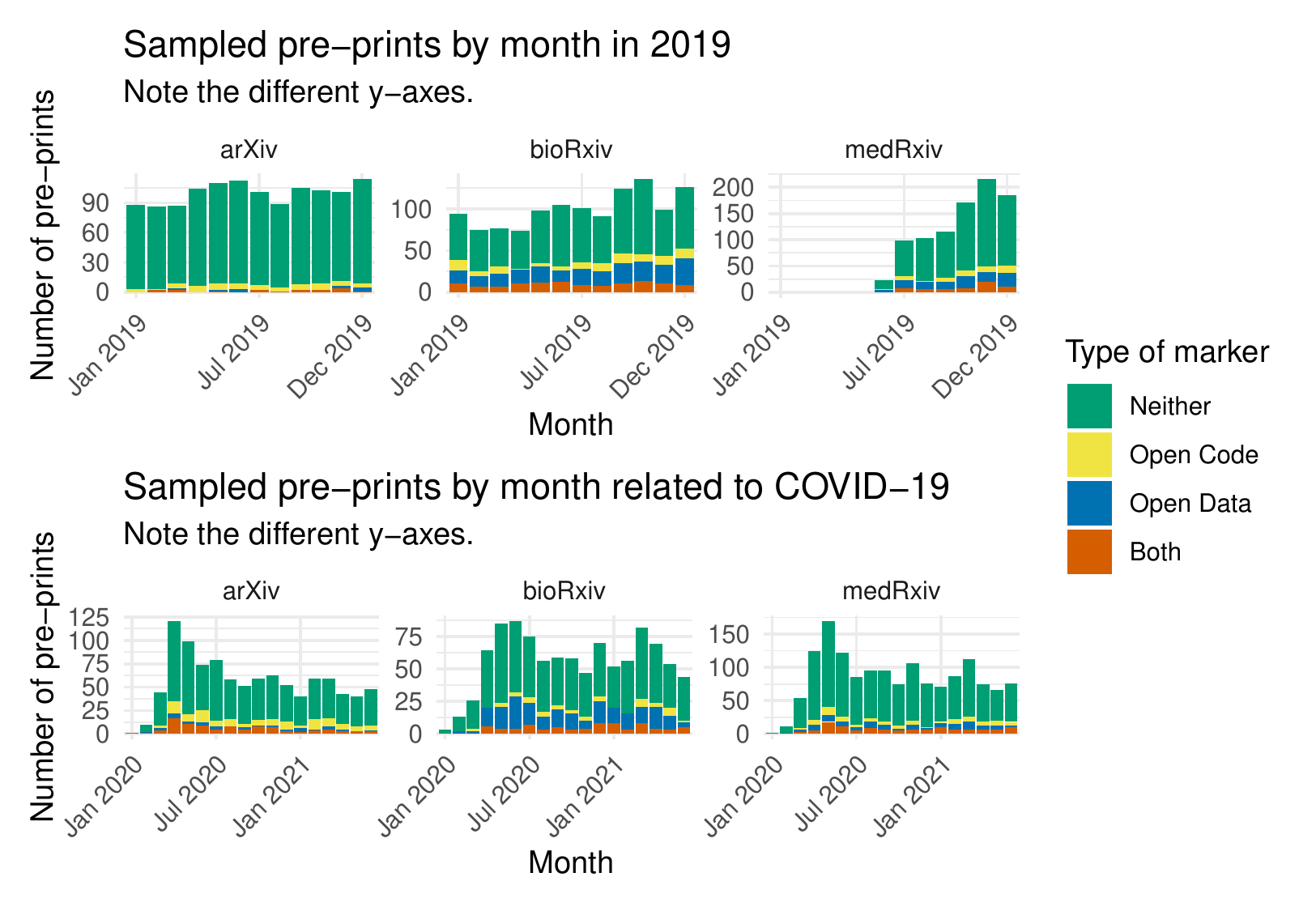} 

}

\caption{Number of sampled pre-prints posted in 2019, compared with those related to COVID-19: from arXiv (n in 2019 = 1,000; n related to COVID = 1,200), bioRxiv (n in 2019 = 1,000; n related to COVID n = 1,200), and medRxiv (n in 2019 = 1,500; n related to COVID n = 913) distinguished by presence of open data or code markers}\label{fig:arxiv-and-monthly-papers-condition-stack}
\end{figure}

The distribution of total sampled pre-prints and sampled pre-prints with open data or code markers roughly follows that of COVID-19-related pre-prints posted in general (Figure \ref{fig:arxiv-and-monthly-papers-condition-stack}). The proportion of pre-prints with open data or code has fluctuated over time but shows no consistent overall increase or decrease throughout the course of the pandemic, nor in conjunction with increases or decreases in the total number of pre-prints posted to any given repository. In our datasets, very few (if any) pre-prints were sampled for the month of January 2020. None of these pre-prints contained open data or open code markers, thus the 0 per cent rate of open data and code for this month across all repositories should be considered an outlier.

It is also important to note that pre-prints posted during the early months of the pandemic were likely using, and reusing, publicly available data sources due to an inability to collect original data within a short timeframe. Additionally, \texttt{oddpub} does not consider `{[}t{]}he reuse of data/code previously published by other researchers' (Riedel, Kip, and Bobrov 2020). A different definition of open data could enable pre-prints that reuse publicly available data to be considered as having their data available for reproducibility purposes.

The proportion of bioRxiv and medRxiv pre-prints lacking both open data and open code are approximately four per cent higher than the corresponding proportions of 2019 pre-prints, suggesting that the analyzed pre-prints from 2019 may contain an overall higher prevalence of open data and code markers than pre-prints concerning COVID-19. Specifically, we found that open data availability in medRxiv pre-prints was significantly associated with a pre-pandemic registration date (\(\chi\)\textsuperscript{2} = 4.8508, p \textless{} 0.005), as was open code availability for bioRxiv pre-prints (\(\chi\)\textsuperscript{2} = 14.491, p \textless{} 0.005). This would suggest that open data and code practices may have suffered in the context of COVID-19, or that it may be something that is backfilled after posting.

On the other hand, the analyzed arXiv COVID-19-related pre-prints contain a higher proportion of open data and code markers overall than their 2019 counterparts, with an increase of 18 per cent in sampled arXiv papers. For these repositories, the presence of both open data and open code markers was significantly associated with registration during the pandemic, suggesting that pre-prints related to COVID-19 in arXiv may have more consistently adhered to open science practices than their pre-pandemic counterparts (\(\chi\)\textsuperscript{2} = 93.124, arXiv open data; \(\chi\)\textsuperscript{2} = 106.88, arXiv open code; \(\chi\)\textsuperscript{2} = 12.303).

\hypertarget{publication-status}{%
\subsection{Publication status}\label{publication-status}}

The proportion of pre-prints that have been published varies by repository (Table \ref{tab:published-sample-summary}). Notably, of all COVID-19-related pre-prints in our dataset, approximately 30 per cent of those posted to bioRxiv and nearly one-third of those posted to medRxiv were published. This is high in comparison to the proportion from arXiv, and although this might suggest that COVID-19-related pre-prints in biomedical fields have received greater attention overall than pre-prints from other fields, our results in Section \ref{pre-pandemic-pre-prints} suggest that this pattern pre-dates the pandemic.

\begin{table}

\caption{\label{tab:published-sample-summary}Counts and proportion of published COVID-19 pre-prints in each repository}
\centering
\begin{tabular}[t]{lrr}
\toprule
Repository & Published pre-prints & Proportion published\\
\midrule
arXiv & 717 & 0.17\\
bioRxiv & 1,230 & 0.30\\
medRxiv & 4,230 & 0.32\\
\bottomrule
\end{tabular}
\end{table}

In Table \ref{tab:open-published-summary} we disaggregate sampled pre-prints by whether there is an indication of publication. We find that the proportion of pre-prints with open data or code markers among those that have been published is roughly the same as pre-prints that have not been published, differing by only a few percentage points.

There is limited literature examining the relationship between data and code availability in manuscripts between the pre-print and publication stages. McGuinness and Sheppard (2021) examine differences in data availability statements between medRxiv pre-prints and their published counterparts. They find that data availability was maintained for most of their sample, varying by journal data sharing policy with greater improvements in openness among manuscripts published in journals mandating data sharing. While limited to medRxiv, the results of McGuinness and Sheppard (2021) align with our own work and provide initial evidence to suggest that data availability is generally maintained or improved between the pre-print and publication stages.

\begin{table}

\caption{\label{tab:open-published-summary}Counts and proportions of open data markers by whether the pre-print was published}
\centering
\begin{tabular}[t]{lrrrrr}
\toprule
Status & Both & Neither & Open code & Open data & Proportion with neither\\
\midrule
\addlinespace[0.3em]
\multicolumn{6}{l}{\textbf{arXiv}}\\
\hspace{1em}Unpublished & 79 & 622 & 100 & 27 & 0.75\\
\hspace{1em}Published & 20 & 130 & 15 & 7 & 0.76\\
\addlinespace[0.3em]
\multicolumn{6}{l}{\textbf{bioRxiv}}\\
\hspace{1em}Unpublished & 55 & 471 & 29 & 145 & 0.67\\
\hspace{1em}Published & 20 & 203 & 15 & 62 & 0.68\\
\addlinespace[0.3em]
\multicolumn{6}{l}{\textbf{medRxiv}}\\
\hspace{1em}Unpublished & 10 & 112 & 8 & 8 & 0.81\\
\hspace{1em}Published & 113 & 1,068 & 81 & 100 & 0.78\\
\bottomrule
\end{tabular}
\end{table}

Our dataset likely imperfectly characterizes publication and does not have the publication details for some papers that were published. And even if it were a perfect record, there is a publication lag (estimated at an average of around 60 days for COVID-19-related pre-prints, although that varies by discipline) that may especially skew the results for pre-prints in the latter portion of our sample (Kwon 2020).

\hypertarget{discussion}{%
\section{Discussion}\label{discussion}}

\hypertarget{on-the-role-of-transparency-and-reproducibility}{%
\subsection{On the role of transparency and reproducibility}\label{on-the-role-of-transparency-and-reproducibility}}

Transparency and reproducibility are hallmarks of quality scientific research due to their relationship with independent verification (Stodden 2020). Open data and open code contribute to both by allowing the scientific community to more easily verify the authenticity of purported scientific discoveries and their supporting evidence. Data sharing also allows others to reuse other researchers' data sets for further analysis or to supplement their own data, contributing to new insights within their field of study.

These factors are especially important in cases where scientific research may quickly and directly impact clinical practice or public policy, such as research on the COVID-19 pandemic. Among many other impacts on the research landscape, COVID-19 has increased the popularity of pre-prints from both a production and consumption standpoint. The number of COVID-19 pre-prints posted to medRxiv increased in the early stages of the pandemic, while non-COVID-19 pre-print numbers were largely as expected. The same trends were apparent in abstracts accessed by medRxiv users, where COVID-19 pre-print abstracts were viewed over 15 times more than non-COVID-19 pre-print abstracts (Fraser et al. 2021). For these reasons, it is important to examine open science standards and reproducibility within pre-print repositories.

Open data is generally accepted to be beneficial to the scientific process and to a paper's reproducibility potential, hence it is concerning that around 75 per cent of pre-prints in our sample contained no open data markers. This concern is slightly mitigated by recognition of challenges in working with biomedical data compared with data in other fields, notably privacy and ethics concerns when working with personal data (Floca 2014). The COVID-19 pandemic has seen open science initiatives, as evidenced by the creation of open data repositories such as the dashboard maintained by the Center for Systems Science and Engineering at Johns Hopkins University (Dong, Du, and Gardner 2020) or the large number of publishers who removed paywalls from published COVID-19 research (Gill 2020). While the intention at the start of the pandemic was that there would be `clear statements regarding the availability of underlying data' (Wellcome Trust 2020) some retractions of work have been based on `unreliable or nonexistent data' (Teixeira da Silva, Bornemann-Cimenti, and Tsigaris 2021).

Open code as an open science marker is context and field-dependent; for instance, not all biomedical research papers will rely on computational methods for their analyses. However, in pre-prints where code comprises a large portion of the methodology or results, posting it openly to repositories like GitHub contributes to a pre-print's potential reproducibility. This is important when computational methods are used to form predictions about emerging situations with limited data or laboratory research, which was the case for modelling studies in the early days of the COVID-19 pandemic. We also see growing concern over the quality and consequences of this sort of research, with bioRxiv no longer allowing purely computational work (Kwon 2020).

The other concern is the adverse selection issue caused by meeting the open science aims of sharing code and data. Authors that share their data and code open their work up to criticism. If authors who make their data and code available make similar mistakes to authors who choose to not publish their data and code, it is more likely that the mistake would not be noticed in the case where data and code were not published. The current system is biased against those who follow best practice. McGuinness and Sheppard (2021) advocate for `(s)trict editorial policies that mandate data sharing,' and other changed norms are needed.

\hypertarget{the-role-of-pre-print-repositories}{%
\subsection{The role of pre-print repositories}\label{the-role-of-pre-print-repositories}}

There has been a large amount of research on COVID-19 (Teixeira da Silva, Tsigaris, and Erfanmanesh 2021). Many concerns have arisen from the rate at which COVID-19 research has been posted and consumed through pre-print repositories, particularly in the early stages of the pandemic (Raynaud et al. 2020). Rushed scientific research has the potential to skip (or at least place less importance on) open science practices, so it may be reasonable to expect a decrease in open data or code markers in the initial few months of the pandemic. We found little relationship between date posted and likelihood of having open data or code markers, with the proportion of pre-prints containing these markers fluctuating from month to month. This suggests that open science practices are more influenced by other factors, perhaps including training, publication bias, or the nature of the pre-print itself. On the other hand, we do not see an overall long-term increase in either open data or open code markers throughout our period of analysis, which we may have expected in the context of the open science movements the pandemic has fostered. Although not pre-print specific, Else (2020) found that overall research output has fluctuated between different fields and topics (namely modelling disease spread, public health, diagnostics and testing, mental health, and hospital mortality) throughout different stages of the pandemic, which may account for some of the fluctuation and overall lack of noticeable trend in our sample.

To emphasize the ongoing need for open data and code in modelling a pandemic, we consider two high profile epidemiological models that emerged in early 2020. Modelling was conducted by Imperial College London (ICL) (Ferguson et al. 2020) and the Institute for Health Metrics and Evaluation (IHME) at the University of Washington (Murray 2020), and both were initially posted to pre-print repositories. The ICL model went on to become the most cited pre-print as of December 2020 (Else 2020), and both had significant influence over policy and public health decisions worldwide (Adam 2020). An independent review of these two models by Jin et al. (2020) found that while code and data were openly available for both, only the ICL model was reproducible due to limited transparency on the underlying methodology of the IHME model. The open-source nature of these models was fundamental to reproduction attempts and is an example of the need for open data and code in COVID-19 research, particularly as pre-prints influence public decision-making.

In the context of the above factors, it was encouraging to find in our analysis that the proportion of pre-prints with open data or code posted to arXiv increased from 7 per cent pre-pandemic to 25 per cent for COVID-19-related pre-prints. This pattern, however, was not observed among the analyzed bioRxiv and medRxiv pre-prints, and may just reflect the nature of COVID-19 pre-prints. With many pre-prints from these repositories still pertaining to epidemiological modelling, one might hope that they should universally be subject to the same analysis as conducted by Jin et al. (2020) as for the examples above, which is made possible by the availability of relevant code and data. Our analysis suggests a need for future investigation and potential overall improvement in open science standards for these types of pre-prints (subject to the data and code considerations already discussed). This need is again emphasized by the new-found speed at which pre-prints may gain public, media, and political attention in the context of the pandemic, particularly those from medRxiv and bioRxiv. One further concern is raised by Teixeira da Silva (2020), who shows that there are pre-prints on those two pre-print servers---medRxiv and bioRxiv---that were withdrawn or retracted with relatively little information about the underlying reason, after gaining substantial media attention.

\hypertarget{the-importance-of-open-data-and-open-code}{%
\subsection{The importance of open data and open code}\label{the-importance-of-open-data-and-open-code}}

Beyond pre-prints, COVID-19 has influenced publication and peer review processes, with timelines for COVID-19 papers being expedited at the expense of longer waits for other scientific research (Else 2020). It is important that open data and code standards be maintained in published work as well. In our sample, published pre-prints contain open data or code markers in similar proportions to their unpublished counterparts, a pattern that was present for pre-prints related to COVID-19 and those posted in 2019. This appears initially to alleviate some concerns over the relationship between open data and publication bias, that is, the potential that journals have favored novel yet less transparent or reproducible papers over those with null results but a high standard of open science practices. However, publication bias is complex, and this result should be approached with caution. Concerns have already been raised through systemic reviews of COVID-19 publications (Raynaud et al. 2020), and oversights in data accessibility have led to high profile retractions of publications in the past; for example, papers from \emph{The Lancet} and the \emph{New England Journal of Medicine} which were withdrawn due to concerns over the private nature of their underlying dataset (Ledford and Noorden 2020). Cabanac, Oikonomidi, and Boutron (2021) show that not all pre-prints are linked to their subsequent peer-reviewed publication, which may further bias our results. Additionally, there is the potential for bias due to older pre-prints having had more time to be published than newer pre-prints. And Oikonomidi et al. (2020) and Bero et al. (2021) show that differences between updated versions of the same pre-print can be substantial; again, this is something that we do not account for and could bias our results.

In all fields of science, increasing access to data and code used for pre-printed or published research is a step in the direction of more transparent, reproducible, and reliable research. The COVID-19 pandemic has created a novel, constantly changing scientific culture that should be navigated with care to uphold standards of scientific practice for both the research community and the safety of the public. Our analysis shows that there is room for improvement in the areas of open data and code availability within COVID-19 pre-print papers on arXiv, bioRxiv, and medRxiv

There is demand for timely research and high frequency results because the pandemic rapidly evolves. Pre-prints are efficient in this role because there is no time spent on peer review. They also allow lesser-known researchers to better disperse their research because of the possibility that fast-tracked peer review may be biased towards established researchers. While there is a clear need for pre-prints, the point remains that they do not go through the peer review process. This question of quality and validity is particularly pertinent in the COVID-19 context because poorly validated results and false information may spread quickly and have real effects. We are not saying that peer review implies that a paper is of a high-quality; we are instead saying that the provision of code and data alongside the pre-print goes some way to allowing others to trust the findings of pre-prints, even though they have not been peer-reviewed. One way this could be encouraged would be for all pre-print repositories to have authors characterize the extent to which they have adopted open science practices as part of their submission, in the same way that is done in SocArXiv. Although those pre-prints that do not adopt these practices should not be rejected from pre-print repositories, greater clarity around this would be useful and might move the state-of-the-art forward.

\hypertarget{weaknesses-and-next-steps}{%
\subsection{Weaknesses and next steps}\label{weaknesses-and-next-steps}}

Future work would expand our analysis to consider the geographic distribution of research and the potential influence of different practices and policies concerning open science. This is important because the epicenter of the pandemic changed throughout the pandemic, which may have implications for our time-based analysis.

A logical next step would be to extend this analysis to additional pre-print servers. We have begun considering samples of pre-pandemic and COVID-19-related pre-prints posted to SocArXiv, a social sciences pre-print server hosted by the Center for Open Science. We validated the ODDPub algorithm against the presence of data links provided by pre-print authors upon submission (available in the pre-print metadata drawn from the Open Science Framework API) and found that the algorithm performed with 52 per cent sensitivity on the 2019 sample and 29 per cent sensitivity for COVID-19-related pre-prints. The high rate of false negatives for open data detection is concerning, and it was decided that the ODDPub algorithm is not suitable for use on pre-prints from this server without modification. A more generalized (or perhaps field-specific) algorithm would be necessary for analysis of open data and code availability in SocArXiv and other more specialized servers. Details of this validation are available in Appendix \ref{oddpub-algorithm-performance-on-socarxiv-pre-prints}.

We recognize that factors beyond open data and code play a large role in the reproducibility of scientific research. Not all pre-prints providing open data or code will be reproducible. Factors such as data documentation, methodological reporting, software choice, and many others all play a role in the reproduction process and should be regarded with just as much gravity when disseminating results.

An important weakness is the potential presence of false negatives in indicators of publication in our dataset. Abdill and Blekhman (2019) estimate that the false-negative rate may be as high as 37.5 per cent for data pulled from the bioRxiv API, meaning analysis of published papers may represent only a fraction of those that have been published. It is unclear to what extent this is the case for other repositories or what bias may exist in the subset of pre-prints for which publication was detected, because it is likely that this process relies on title-based text matching (Abdill and Blekhman 2019). It is also likely that some of our more recent sampled pre-prints will be published in the future which we could not account for at the time of our data collection.

Our paper depends on search responses from the various repositories, which are based on our selection of keywords. Our selection of keywords is not exhaustive, for instance, perhaps `the pandemic' could result in additional papers. Future work could make this keyword approach more systematic, for instance following King, Lam, and Roberts (2017).

We also recognize that this analysis relies heavily on text-based analysis which was not verified directly in most cases and may lead to higher levels of uncertainty. The \texttt{oddpub} package was built to analyze biomedical publications and it may be that some of the differences that we find between repositories are due to this. We also note that the ODDPub algorithm is relatively narrow in its definition of ``open,'' excluding data that is available via registration or in some other restricted form. Considering a broader definition of openness, either through using a less restrictive algorithm or through manual verification, would likely produce different results particularly for pre-prints using clinical data. Future work could take smaller sub-samples to validate factors like publication status, paper topic, and open code and data status, beyond the approaches we used here.

\newpage

\hypertarget{appendix-appendix}{%
\appendix}

\hypertarget{oddpub-algorithm-performance-on-arxiv-pre-prints}{%
\section{ODDPub algorithm performance on arXiv pre-prints}\label{oddpub-algorithm-performance-on-arxiv-pre-prints}}

We verified the accuracy of the ODDPub algorithm on a subset of our analyzed pre-prints from 2019 from arXiv. We took a simple random sample of 100 papers. In the original validation process, the annotators stratified by detection status prior to sampling to ensure relatively high representation of papers where open data or code was detected. Since the major concern for our manual verification is potential false negatives, this skewed representation was unnecessary. Open data and code status were verified first via the ``Code \& Data'' tab on each pre-print's page on the arXiv website, then by checking for an explicit data availability section within the pre-print PDF, and finally by manually checking the body of the paper using keyword searches. Results were recorded manually in Excel. This mimics the procedure outlined for the original validation of ODDPub (Riedel, Kip, and Bobrov 2020).

Many of the pre-prints in arXiv did not use data or code, namely those from pure mathematics and physics. There were also several that reused other publicly or privately available data sets, and regardless of whether or not they were shared alongside the paper, these do not count as open data according to the standards outlined by the original authors of the ODDPub algorithm (Riedel, Kip, and Bobrov 2020). Algorithmic performance is specified in Table \ref{tab:arxiv-accuracy-data-2019} and Table \ref{tab:arxiv-accuracy-code-2019}.

\begin{table}[!h]

\caption{\label{tab:arxiv-confusion-data-2019}ODDPub predictions for open data compared with manual check, arXiv sample}
\centering
\begin{tabular}[t]{lrr}
\toprule
Predicted & Data available & No data available\\
\midrule
Open data detected & 2 & 2\\
No open data detected & 1 & 95\\
\bottomrule
\end{tabular}
\end{table}

\begin{table}[!h]

\caption{\label{tab:arxiv-accuracy-data-2019}ODDPub prediction accuracy, open data, arXiv sample}
\centering
\begin{tabular}[t]{lr}
\toprule
Metric & Value\\
\midrule
Accuracy & 0.97\\
Sensitivity & 0.67\\
Specificity & 0.98\\
\bottomrule
\end{tabular}
\end{table}

\begin{table}[!h]

\caption{\label{tab:arxiv-confusion-code-2019}ODDPub predictions for open code compared with manual check, arXiv sample}
\centering
\begin{tabular}[t]{lrr}
\toprule
Predicted & Code available & No code available\\
\midrule
Open code detected & 3 & 2\\
No open code detected & 2 & 93\\
\bottomrule
\end{tabular}
\end{table}

\begin{table}[!h]

\caption{\label{tab:arxiv-accuracy-code-2019}ODDPub prediction accuracy, open code, arXiv sample}
\centering
\begin{tabular}[t]{lr}
\toprule
Metric & Value\\
\midrule
Accuracy & 0.96\\
Sensitivity & 0.60\\
Specificity & 0.98\\
\bottomrule
\end{tabular}
\end{table}

\newpage

\hypertarget{supporting-tables}{%
\section{Supporting tables}\label{supporting-tables}}

\begin{table}[!h]

\caption{\label{tab:2019-open-results}Count and proportions of open data and code markers by pre-print repository in 2019 sample}
\centering
\begin{tabular}[t]{lrr}
\toprule
Markers & Count & Proportion of total\\
\midrule
\addlinespace[0.3em]
\multicolumn{3}{l}{\textbf{Total}}\\
\hspace{1em}Neither & 2,550 & 0.77\\
\hspace{1em}Open code & 212 & 0.06\\
\hspace{1em}Open data & 361 & 0.11\\
\hspace{1em}Both & 190 & 0.06\\
\addlinespace[0.3em]
\multicolumn{3}{l}{\textbf{arXiv}}\\
\hspace{1em}Neither & 1,112 & 0.93\\
\hspace{1em}Open code & 59 & 0.05\\
\hspace{1em}Open data & 12 & 0.01\\
\hspace{1em}Both & 17 & 0.01\\
\addlinespace[0.3em]
\multicolumn{3}{l}{\textbf{bioRxiv}}\\
\hspace{1em}Neither & 752 & 0.63\\
\hspace{1em}Open code & 98 & 0.08\\
\hspace{1em}Open data & 234 & 0.20\\
\hspace{1em}Both & 116 & 0.10\\
\addlinespace[0.3em]
\multicolumn{3}{l}{\textbf{medRxiv}}\\
\hspace{1em}Neither & 686 & 0.75\\
\hspace{1em}Open code & 55 & 0.06\\
\hspace{1em}Open data & 115 & 0.13\\
\hspace{1em}Both & 57 & 0.06\\
\bottomrule
\end{tabular}
\end{table}

\begin{table}[!h]

\caption{\label{tab:summarycounts-repositories}Count and proportions of open data and code markers by pre-print repository}
\centering
\begin{tabular}[t]{lrr}
\toprule
Markers & Count & Proportion of total\\
\midrule
\addlinespace[0.3em]
\multicolumn{3}{l}{\textbf{Total}}\\
\hspace{1em}Neither & 2,606 & 0.74\\
\hspace{1em}Open code & 248 & 0.07\\
\hspace{1em}Open data & 349 & 0.10\\
\hspace{1em}Both & 297 & 0.08\\
\addlinespace[0.3em]
\multicolumn{3}{l}{\textbf{arXiv}}\\
\hspace{1em}Neither & 752 & 0.75\\
\hspace{1em}Open code & 115 & 0.12\\
\hspace{1em}Open data & 34 & 0.03\\
\hspace{1em}Both & 99 & 0.10\\
\addlinespace[0.3em]
\multicolumn{3}{l}{\textbf{bioRxiv}}\\
\hspace{1em}Neither & 674 & 0.67\\
\hspace{1em}Open code & 44 & 0.04\\
\hspace{1em}Open data & 207 & 0.21\\
\hspace{1em}Both & 75 & 0.07\\
\addlinespace[0.3em]
\multicolumn{3}{l}{\textbf{medRxiv}}\\
\hspace{1em}Neither & 1,180 & 0.79\\
\hspace{1em}Open code & 89 & 0.06\\
\hspace{1em}Open data & 108 & 0.07\\
\hspace{1em}Both & 123 & 0.08\\
\bottomrule
\end{tabular}
\end{table}

\begin{table}[!h]

\caption{\label{tab:propbyserverandcategory}Number and proportion of all preprints in each category, for each preprint server}
\centering
\begin{tabular}[t]{lrr}
\toprule
Markers & Count & Proportion of total\\
\midrule
\addlinespace[0.3em]
\multicolumn{3}{l}{\textbf{arXiv}}\\
\hspace{1em}Both & 99 & 0.10\\
\hspace{1em}Neither & 752 & 0.75\\
\hspace{1em}Open Code & 115 & 0.12\\
\hspace{1em}Open Data & 34 & 0.03\\
\addlinespace[0.3em]
\multicolumn{3}{l}{\textbf{bioRxiv}}\\
\hspace{1em}Both & 75 & 0.07\\
\hspace{1em}Neither & 674 & 0.67\\
\hspace{1em}Open Code & 44 & 0.04\\
\hspace{1em}Open Data & 207 & 0.21\\
\addlinespace[0.3em]
\multicolumn{3}{l}{\textbf{medRxiv}}\\
\hspace{1em}Both & 123 & 0.08\\
\hspace{1em}Neither & 1,180 & 0.79\\
\hspace{1em}Open Code & 89 & 0.06\\
\hspace{1em}Open Data & 108 & 0.07\\
\bottomrule
\end{tabular}
\end{table}

\newpage

\hypertarget{oddpub-algorithm-performance-on-socarxiv-pre-prints}{%
\section{ODDPub algorithm performance on SocArXiv pre-prints}\label{oddpub-algorithm-performance-on-socarxiv-pre-prints}}

SocArXiv allows authors to input a link to their data source/repository upon submission of a pre-print. This link can then be accessed via the API metadata. The presence of a data link was used as an indicator that a pre-print provides open data for the purposes of validating the ODDPub algorithm. When available, a data link is stored under the variable name ``attributes.data\_links.'' The data was manipulated using functions from the R package \texttt{tidyverse} (Wickham et al. 2019) to create a binary variable indicating data availability or lack thereof. We assume ``attributes.data\_links'' to indicate the true availability of data for the purposes of validating the ODDPub algorithm. It is possible, however, that some authors failed to indicate their data availability in the proper field upon posting to SocArXiv, and thus some of the false positive may in fact be true positives.

Against the data availability indicated by pre-print authors in our 2019 sample, the ODDPub algorithm performed with an accuracy of 93 percent, a sensitivity of 52 percent, and a specificity of 94 percent. In our 2020 and 2021 sample, the algorithm performed with an accuracy of 79 percent, a sensitivity of 29 percent, and a specificity of 92 percent. Specific predictions are broken down in Table \ref{tab:confusion-matrix-2019} and Table \ref{tab:confusion-matrix-2020}.

It is unclear the precise inclusion criteria for data submitted to the data link field. It is possible that some of the links provided lead to data sets that are publicly available for reuse, which would not constitute ``open data'' by the ODDPub algorithm's definition, in which case the accuracy could potentially be higher in reality than 93 percent and 79 percent in the samples considered.

\begin{table}[!h]

\caption{\label{tab:confusion-matrix-2019}ODDPub predictions for open data compared with data links provided by authors, 2019 SocArXiv sample}
\centering
\begin{tabular}[t]{lrr}
\toprule
ODDPub algorithm & Data linked & No data linked\\
\midrule
Open data detected & 11 & 72\\
No open data detected & 10 & 1107\\
\bottomrule
\end{tabular}
\end{table}

\begin{table}

\caption{\label{tab:accuracy-2019}ODDPub prediction accuracy, open data, 2019 SocArXiv sample}
\centering
\begin{tabular}[t]{lr}
\toprule
Metric & Value\\
\midrule
Accuracy & 0.93\\
Sensitivity & 0.52\\
Specificity & 0.94\\
\bottomrule
\end{tabular}
\end{table}

\begin{table}[!h]

\caption{\label{tab:confusion-matrix-2020}ODDPub predictions for open data compared with data links provided by authors, COVID-19-related SocArXiv pre-prints sample}
\centering
\begin{tabular}[t]{lrr}
\toprule
ODDPub algorithm & Data linked & No data linked\\
\midrule
Open data detected & 28 & 31\\
No open data detected & 69 & 350\\
\bottomrule
\end{tabular}
\end{table}

\begin{table}[!h]

\caption{\label{tab:accuracy-2020}ODDPub prediction accuracy, open data, COVID-19-related SocArXiv pre-prints sample}
\centering
\begin{tabular}[t]{lr}
\toprule
Metric & Value\\
\midrule
Accuracy & 0.79\\
Sensitivity & 0.29\\
Specificity & 0.92\\
\bottomrule
\end{tabular}
\end{table}

\newpage

\hypertarget{references}{%
\section*{References}\label{references}}
\addcontentsline{toc}{section}{References}

\hypertarget{refs}{}
\begin{CSLReferences}{1}{0}
\leavevmode\hypertarget{ref-abdill2019}{}%
Abdill, Richard J, and Ran Blekhman. 2019. {``Meta-Research: Tracking the Popularity and Outcomes of All bioRxiv Preprints.''} Edited by Emma Pewsey, Peter Rodgers, and Casey S Greene. \emph{eLife} 8 (April): e45133. \url{https://doi.org/10.7554/eLife.45133}.

\leavevmode\hypertarget{ref-adam2020}{}%
Adam, David. 2020. {``Special Report: The Simulations Driving the World's Response to COVID-19.''} \emph{Nature} 580: 316--18. \url{https://doi.org/10.1038/d41586-020-01003-6}.

\leavevmode\hypertarget{ref-barba2018terminologies}{}%
Barba, Lorena A. 2018. {``Terminologies for Reproducible Research.''} \url{http://arxiv.org/abs/1802.03311}.

\leavevmode\hypertarget{ref-Beroe051821}{}%
Bero, Lisa, Rosa Lawrence, Louis Leslie, Kellia Chiu, Sally McDonald, Matthew J Page, Quinn Grundy, et al. 2021. {``Cross-Sectional Study of Preprints and Final Journal Publications from COVID-19 Studies: Discrepancies in Results Reporting and Spin in Interpretation.''} \emph{BMJ Open} 11 (7). \url{https://doi.org/10.1136/bmjopen-2021-051821}.

\leavevmode\hypertarget{ref-cabanac2021day}{}%
Cabanac, Guillaume, Theodora Oikonomidi, and Isabelle Boutron. 2021. {``Day-to-Day Discovery of Preprint--Publication Links.''} \emph{Scientometrics} 126 (6): 5285--5304.

\leavevmode\hypertarget{ref-cacioppo2015social}{}%
Cacioppo, John T, Robert M Kaplan, Jon A Krosnick, James L Olds, and Heather Dean. 2015. {``Social, Behavioral, and Economic Sciences Perspectives on Robust and Reliable Science.''} \emph{Report of the Subcommittee on Replicability in Science Advisory Committee to the National Science Foundation Directorate for Social, Behavioral, and Economic Sciences}.

\leavevmode\hypertarget{ref-collins2021reproducibility}{}%
Collins, Annie, and Rohan Alexander. 2021. {``Reproducibility of COVID-19 Pre-Prints.''} \url{http://arxiv.org/abs/2107.10724}.

\leavevmode\hypertarget{ref-citeCSSE}{}%
Dong, Ensheng, Hongru Du, and Lauren Gardner. 2020. {``An Interactive Web-Based Dashboard to Track COVID-19 in Real Time.''} \emph{The Lancet Infectious Diseases} 20: 533--34. \url{https://doi.org/10.1016/S1473-3099(20)30120-1}.

\leavevmode\hypertarget{ref-else2020}{}%
Else, Holly. 2020. {``How a Torrent of COVID Science Changed Research Publishing --- in Seven Charts.''} \emph{Nature} 588: 553. \url{https://doi.org/10.1038/d41586-020-03564-y}.

\leavevmode\hypertarget{ref-ferguson2020}{}%
Ferguson, Neil M, Daniel Laydon, Gemma Nedjati-Gilani, Natsuko Imai, Kylie Ainslie, Marc Baguelin, Sangeeta Bhatia, et al. 2020. {``Report 9: Impact of Non-Pharmaceutical Interventions (NPIs) to Reduce COVID-19 Mortality and Healthcare Demand.''} \url{https://doi.org/10.25561/77482}.

\leavevmode\hypertarget{ref-Floca2014}{}%
Floca, Ralf. 2014. {``Challenges of Open Data in Medical Research.''} In \emph{Opening Science: The Evolving Guide on How the Internet Is Changing Research, Collaboration and Scholarly Publishing}, edited by Sönke Bartling and Sascha Friesike, 297--307. Cham: Springer International Publishing. \url{https://doi.org/10.1007/978-3-319-00026-8_22}.

\leavevmode\hypertarget{ref-citeFraser}{}%
Fraser, Nicholas, Liam Brierley, Gautam Dey, Jessica K Polka, Máté Pálfy, Federico Nanni, and Jonathon Alexis Coates. 2021. {``Preprinting the COVID-19 Pandemic.''} \emph{bioRxiv}. \url{https://doi.org/10.1101/2020.05.22.111294}.

\leavevmode\hypertarget{ref-gill2020}{}%
Gill, David. 2020. {``Immediate Free Access to Research: The Scholarly Response to COVID-19.''} \url{https://www.lib.sfu.ca/help/publish/scholarly-publishing/radical-access/scholarly-covid19}.

\leavevmode\hypertarget{ref-jin2020}{}%
Jin, Jin, Neha Agarwala, Prosenjit Kundu, Yi Wang, Ruzhang Zhao, and Nilanjan Chatterjee. 2020. {``Transparency, Reproducibility, and Validation of COVID-19 Projection Models.''} \url{https://www.jhsph.edu/covid-19/articles/transparency-reproducibility-and-validation-of-covid-19-projection-models.html}.

\leavevmode\hypertarget{ref-king2017computer}{}%
King, Gary, Patrick Lam, and Margaret E Roberts. 2017. {``Computer-Assisted Keyword and Document Set Discovery from Unstructured Text.''} \emph{American Journal of Political Science} 61 (4): 971--88.

\leavevmode\hypertarget{ref-kwon2020}{}%
Kwon, Diana. 2020. {``How Swamped Preprint Servers Are Blocking Bad Coronavirus Research.''} \emph{Nature} 580 (May): 130--31. \url{https://doi.org/10.1038/d41586-020-01394-6}.

\leavevmode\hypertarget{ref-ledford2020}{}%
Ledford, Heidi, and Richard Van Noorden. 2020. {``High-Profile Coronavirus Retractions Raise Concerns about Data Oversight.''} \emph{Nature} 582: 160. \url{https://doi.org/10.1038/d41586-020-01695-w}.

\leavevmode\hypertarget{ref-mcguinness2021descriptive}{}%
McGuinness, Luke A, and Athena L Sheppard. 2021. {``A Descriptive Analysis of the Data Availability Statements Accompanying medRxiv Preprints and a Comparison with Their Published Counterparts.''} \emph{PloS One} 16 (5): e0250887.

\leavevmode\hypertarget{ref-murray2020}{}%
Murray, Christopher JL. 2020. {``Forecasting the Impact of the First Wave of the COVID-19 Pandemic on Hospital Demand and Deaths for the USA and European Economic Area Countries.''} \emph{medRxiv}. \url{https://doi.org/10.1101/2020.04.21.20074732}.

\leavevmode\hypertarget{ref-NAP25303}{}%
National Academies of Sciences, Engineering, and Medicine. 2019. \emph{Reproducibility and Replicability in Science}. Washington, DC: The National Academies Press. \url{https://doi.org/10.17226/25303}.

\leavevmode\hypertarget{ref-oikonomidi2020changes}{}%
Oikonomidi, Theodora, Isabelle Boutron, Olivier Pierre, Guillaume Cabanac, and Philippe Ravaud. 2020. {``Changes in Evidence for Studies Assessing Interventions for COVID-19 Reported in Preprints: Meta-Research Study.''} \emph{BMC Medicine} 18 (1): 1--10.

\leavevmode\hypertarget{ref-citeR}{}%
R Core Team. 2020. \emph{{R: A Language and Environment for Statistical Computing}}. Vienna, Austria: R Foundation for Statistical Computing. \url{https://www.R-project.org/}.

\leavevmode\hypertarget{ref-raynaud2021}{}%
Raynaud, Marc, Huanxi Zhang, Kevin Louis, Valentin Goutaudier, Jiali Wang, Quentin Dubourg, Yongcheng Wei, et al. 2020. {``COVID-19-Related Medical Research: A Meta-Research and Critical Appraisal.''} \emph{BMC Medical Research Methodology} 21. \url{https://doi.org/10.1186/s12874-020-01190-w}.

\leavevmode\hypertarget{ref-citeODDPubpackage}{}%
Riedel, Nico. 2019. \emph{Oddpub: Detection of Open Data \& Open Code Statements in Biomedical Publications}. \url{https://github.com/quest-bih/oddpub}.

\leavevmode\hypertarget{ref-citePDDPub}{}%
Riedel, Nico, Miriam Kip, and Evgeny Bobrov. 2020. {``ODDPub -- a Text-Mining Algorithm to Detect Data Sharing in Biomedical Publications.''} \emph{Data Science Journal} 19 (1): 42. \url{https://doi.org/10.5334/dsj-2020-042}.

\leavevmode\hypertarget{ref-Stodden2020Theme}{}%
Stodden, Victoria. 2020. {``Theme Editor's Introduction to Reproducibility and Replicability in Science.''} \emph{Harvard Data Science Review} 2 (4). \url{https://doi.org/10.1162/99608f92.c46a02d4}.

\leavevmode\hypertarget{ref-da2020silently}{}%
Teixeira da Silva, Jaime A. 2020. {``Silently Withdrawn or Retracted Preprints Related to Covid-19 Are a Scholarly Threat and a Potential Public Health Risk: Theoretical Arguments and Suggested Recommendations.''} \emph{Online Information Review}.

\leavevmode\hypertarget{ref-da2021optimizing}{}%
Teixeira da Silva, Jaime A, Helmar Bornemann-Cimenti, and Panagiotis Tsigaris. 2021. {``Optimizing Peer Review to Minimize the Risk of Retracting COVID-19-Related Literature.''} \emph{Medicine, Health Care and Philosophy} 24 (1): 21--26.

\leavevmode\hypertarget{ref-da2021publishing}{}%
Teixeira da Silva, Jaime A, Panagiotis Tsigaris, and Mohammadamin Erfanmanesh. 2021. {``Publishing Volumes in Major Databases Related to Covid-19.''} \emph{Scientometrics} 126 (1): 831--42.

\leavevmode\hypertarget{ref-weissgerber2021automated}{}%
Weissgerber, Tracey, Nico Riedel, Halil Kilicoglu, Cyril Labbé, Peter Eckmann, Gerben Ter Riet, Jennifer Byrne, et al. 2021. {``Automated Screening of COVID-19 Preprints: Can We Help Authors to Improve Transparency and Reproducibility?''} \emph{Nature Medicine} 27 (1): 6--7.

\leavevmode\hypertarget{ref-citeWellcome}{}%
Wellcome Trust. 2020. \emph{Sharing Research Data and Findings Relevant to the Novel Coronavirus (COVID-19) Outbreak}. Vienna, Austria: Wellcome Trust. \url{https://wellcome.org/coronavirus-covid-19/open-data}.

\leavevmode\hypertarget{ref-citetidyverse}{}%
Wickham, Hadley, Mara Averick, Jennifer Bryan, Winston Chang, Lucy D'Agostino McGowan, Romain François, Garrett Grolemund, et al. 2019. {``Welcome to the {tidyverse}.''} \emph{Journal of Open Source Software} 4 (43): 1686. \url{https://doi.org/10.21105/joss.01686}.

\leavevmode\hypertarget{ref-citeWHOtimeline}{}%
World Health Organization. 2020. {``WHO Director-General's Opening Remarks at the Media Briefing on COVID-19 - 11 March 2020.''} \url{https://www.who.int/director-general/speeches/detail/who-director-general-s-opening-remarks-at-the-media-briefing-on-covid-19---11-march-2020}.

\end{CSLReferences}

\end{document}